\documentclass{article}
\usepackage{ijcai25}

\usepackage{times}
\usepackage{soul}
\usepackage{url}
\usepackage[hidelinks]{hyperref}
\usepackage[utf8]{inputenc}
\usepackage[small]{caption}
\usepackage{graphicx}
\usepackage{amsmath}
\usepackage{amssymb}
\usepackage{amsthm}
\usepackage{booktabs}
\usepackage{algorithm}
\usepackage{algorithmic}

\usepackage{braket}
\usepackage{svg}
\usepackage{stfloats}

\title{Multiple Embeddings for Quantum Machine Learning }

\author{
Siyu Han$^1$
\and
Lihan Jia$^2$\and
Lanzhe Guo$^{3}$
\affiliations
$^{1, 2, 3}$Nanjing University
\emails
\{hansy, jialh\}@lamda.nju.edu.cn,
guolz@lamda.nju.edu.cn,
}
\newcommand{\bs}[1]{\boldsymbol{#1}}

\begin{document}

\maketitle

\begin{abstract}
    This work focuses on the limitations about the insufficient fitting capability of current quantum machine learning methods, which results from the over-reliance on a single data embedding strategy. We propose a novel quantum machine learning framework that integrates multiple quantum data embedding strategies, allowing the model to fully exploit the diversity of quantum computing when processing various datasets. Experimental results validate the effectiveness of the proposed framework, demonstrating significant improvements over existing state-of-the-art methods and achieving superior performance in practical applications.
\end{abstract}

\section{Introduction}

Since its introduction by Richard Feynman in the 1980s, quantum computing has demonstrated unique advantages in simulating quantum systems. With the development of Shor's algorithm and Grover's search algorithm, quantum computing has shown performance that surpasses classical computers in areas such as cryptography and search problems. In 2019, Google announced that their quantum computer, Sycamore, achieved "quantum supremacy," meaning that for certain specific tasks, the performance of quantum computers exceeded that of the most powerful classical computers. This milestone has garnered wider academic attention to the field.

With the transition of quantum computers from theoretical concepts to practical systems, an increasing number of researchers have realized the advantages of quantum computers over classical computers in handling complex computational problems. As a result, research in quantum machine learning, which involves performing machine learning on quantum computers, has seen rapid growth in recent years.

%量子计算自从1980年代由理查德·费曼（Richard Feynman）提出以来，便在模拟量子系统上的独特优势。而随着Shor算法和Grover搜索算法的提出，量子计算在密码学和搜索等问题上表现出了经典计算机难以企及的优势。2019年，Google宣布他们的量子计算机Sycamore实现了“量子霸权”，即在某些特定任务上量子计算机的性能超过了最强大的经典计算机。使得这个领域得到学术界了更广泛的关注。
%随着量子计算机由理论构想到实际系统的过渡，越来越多的研究人员意识到了量子计算机在处理复杂计算问题方面相较于经典计算机的优势，因此在量子计算机上进行机器学习，也就是量子机器学习的相关研究在近几年迎来了蓬勃的发展。
Despite its theoretical soundness, the practical application of the quantum machine learning models demonstrates suboptimal performance on certain datasets, for example, linearly separable datasets\cite{bowles2024betterclassicalsubtleart}. The influence of data encoding on decision boundaries remains significant, which reveals that the generalization capability of the quantum machine learning model still has substantial room for improvement.

The lack of generalization capability originates from the model's reliance on a single data encoding, which, similar to ANNs, limits the model's ability to extract features effectively when dealing with specific datasets. Consequently, this limitation impacts the overall generalization capability, and this issue cannot be resolved merely by replacing the data encoding. Therefore, it is imperative to propose a method that integrates multiple data embeddings and leverages their combined strengths. This is precisely the goal of our work. Our proposed approach not only maintains performance on existing datasets but also achieves up to 20\% performance improvement on certain datasets.

The contributions of the article are as follows:

\begin{enumerate}
    \item We investigate a novel problem of why quantum machine learning models lack the capability to generalize linear separable datasets.
    \item We propose a new network framework which can integrate multiple data embeddings based on the analysis.
    \item We evaluate the framework on the benchmarks, which shows that the proposed method significantly outperforms state-of-the-art quantum machine learning models.
\end{enumerate}

\begin{figure*}[!t]
    \centering
    \includegraphics[width=0.5\linewidth]{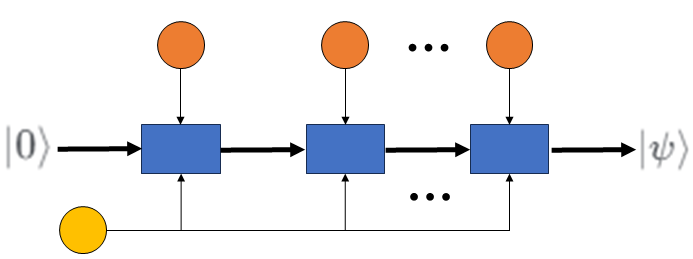}
    \caption{Visual illustration of Data Reuploading Model}
    \label{fig:data-reuplod}
\end{figure*}

\section{Related Works}
 Machine learning has witnessed significant evolution recent years. Early works in ANNs, inspired by biological neurons, focused on creating models capable of pattern recognition and classification tasks \cite{10.1037/h0042519}. However, it was not until the advent of deep learning that neural networks reached their full potential. Deep learning models, particularly deep neural networks (DNNs), enabled the automatic extraction of hierarchical features from data, leading to breakthroughs in tasks such as image recognition, speech processing, and natural language understanding \cite{lecun_deep_2015}. CNNs, a specialized type of neural network, revolutionized computer vision by efficiently processing grid-like data, such as images, and achieving state-of-the-art performance in various visual recognition benchmarks \cite{NIPS2012_c399862d}. These advances, combined with increased computational power and large-scale datasets, have made deep learning the dominant approach in modern machine learning research and applications.

Quantum machine learning has seen extensive research in recent years, with many quantum versions of traditional machine learning methods being developed\cite{biamonte_quantum_2017} \cite{e25020287} \cite{cerezo_challenges_2022} \cite{tychola_quantum_2023}. Researchers have studied quantum machine learning models from various directions. With regard to training data, the article \cite{ghobadi_power_2019} explores the impact of quantum data on both quantum machine learning and classical machine learning. \cite{caro_generalization_2022} optimized quantum machine learning models to reduce the amount of data required for training. The relationship between quantum machine learning and classical machine learning has become a popular research topic. \cite{schreiber_classical_2023} introduces the concept of surrogate models in machine learning, where classical machine learning methods are used to simulate quantum machine learning models, thereby reducing the overuse of qubits. Furthermore, the shadow model\cite{jerbi_shadows_2024} guided by \cite{huang_predicting_2020} in the field of quantum mechanics provides an insightful approach. This model involves performing several observations on qubits and reconstructing the qubits in a classical computer, which offers a valuable perspective for further research.

In general, quantum machine learning models are divided into implicit kernel methods, such as quantum support vector machines\cite{havlicek_supervised_2019}, and explicit quantum models based on variational quantum circuits\cite{jerbi_quantum_2023} \cite{jerbi_shadows_2024}.

Explicit quantum models, as described in \cite{jerbi_shadows_2024}, typically consist of the following components: data encoding, quantum circuits\cite{guala_practical_2023}, and measurements.In terms of data encoding\cite{Caro_2021}, various encoding schemes have been developed to address different machine learning tasks\cite{larose_robust_2020}, with common approaches including amplitude encoding, phase encoding, and QAOA encoding\cite{lloyd2020quantumembeddingsmachinelearning}, among others. For quantum circuits, models have been developed that combine classical neural networks with quantum neural networks (QCNN)\cite{Cong_2019}, as well as data reuploading models\cite{perez-salinas_data_2020} based on artificial neural networks (ANNs)\cite{alzubaidi_review_2021}. In the measurement aspect, some works aim to establish shadow models through quantum measurements\cite{huang_predicting_2020}, hoping to leverage classical machine learning models to achieve quantum advantages on specific tasks. Additionally, there is also research into the relationship between quantum machine learning and classical machine learning\cite{huang_power_2021}.

However, quantum machine learning has faced challenges when applied to practical use, especially in classification tasks\cite{bowles2024betterclassicalsubtleart} \cite{kavitha_quantum_2024}. Due to the inherent characteristics of quantum machine learning—namely, the data encoding stage that transforms classical data into quantum bits—different encoding methods\cite{schuld_2021} result in varying representations of data within the quantum bits. This, in turn, affects the decision boundaries of each model. \cite{larose_robust_2020}Therefore, it can be said that while data encoding makes it possible for quantum machine learning to handle classical machine learning tasks, it also limits the ability of quantum machine learning to effectively handle classification tasks.

One of the most representative attempts in this area is the Data Reuploading Classifier, which draws inspiration from artificial neural networks (ANNs)\cite{perez-salinas_data_2020}. It re-encodes the data and reloads it into the quantum machine learning model, treating each encoding as a "neuron." The universal approximation theorem ensures its strong generalization capabilities, a fact that has been validated in relevant benchmarks.\cite{bowles2024betterclassicalsubtleart}

This paper is organized as follows: The first two Sections provide a brief overview of the background and related works in quantum machine learning. In Section 3, we introduce the current state-of-the-art method in quantum machine learning, the data reuploading model. Section 4 presents the details of our proposed framework. Section 5 showcases the experimental results, followed by discussions in Section 6.
%量子机器学习近几年已经有了很多人去研究这个领域。许多传统的机器学习方法的量子形式被开发出来。分为以量子支持向量机为代表的隐式核方法，以及由变分量子线路构成的显式量子模型。[Shadows of quantum]显式量子模型一般由以下几个部分组成：数据编码，量子电路以及测量部分[Circuit-centric quantum classiers]。而针对以下这几个组成部分的优化一直在被研究者们广泛进行着。在数据编码方面，有多种多样的编码方式被开发出来以应对各种机器学习任务，其中常见的有振幅编码，相位编码，QAOA编码等。在量子电路方面，已经有将神经网络与量子神经网络进行结合的模型（QCNN），也有参考ANN所开发的数据重载模型。在测量方面，有一些工作希望通过对量子模型的测量建立阴影模型，从而希望利用经典机器学习模型在特定任务上实现量子优势。除此以外，也有一些量子机器学习和经典机器学习之间关系的探究。[power of quantum ]

%然而一直以来，量子机器学习在投入实际使用时，尤其是分类任务，由于量子机器学习的本身的特色，即将经典数据转换为量子比特的数据编码阶段，不同的数据编码会导致数据在量子比特中不同的存在形式，进而影响每一个模型的决策界。因此可以说数据编码在让量子机器学习处理经典机器学习任务成为可能的同时，也限制了量子机器学习在处理分类任务的能力。

%在这一方面做出尝试的相关工作中，最有代表性的工作就是DataReuploading Classifier，它借鉴ANN，通过重复编码的方式将数据重载入量子机器学习模型中，将每一次的编码视为一个神经元，进而万能逼近定理可以保证其强大的泛化能力，这也在相关的benchmark中得到了验证[benchmark]. 然而，数据编码导致的复杂分类任务的能力限制仍然存在于数据重载模型当中，尤其是当层数偏少的情况下，此时万能逼近定理的条件不再能被满足，数据编码对决策界的影响变得不可忽视。这与ANN类似——因为其简单的结构导致缺乏能够高效提取有效特征的模块。

%在经典机器学习中，这个问题在CNN通过设计好的多种神经网络模块从而得到了改善和解决。因此，由CNN启发，如果将每一种数据编码视为一个独特的模块，并通过某种方式将其组合起来，就能苟实现对于多种特征的提取，改善量子机器学习模型在分类任务上提取复杂特征的能力。

%本文依照以下结构展开：在第二节我们会首先介绍量子机器学习和神经网络的基础知识；在第三节我们会介绍我们所提出的结构；第四节是进行实验结果的展示；最后进行总结和讨论。

\section{Preliminary}
\subsection{Data Reuploading Classifier}
The Data Reuploading Classifier, proposed by Pérez-Salinas et al. (2020), was initially designed to create a universal quantum classifier that uses minimal quantum resources. The core idea of the model addresses the issue of limited computational space for a single quantum bit (only two degrees of freedom). To approximate complex classification functions, the model repeatedly "reuploads" data within the quantum circuit. This data re-uploading bypasses the no-cloning theorem in quantum computing (which prohibits directly copying quantum data) by reintroducing classical data into the quantum circuit at multiple layers, achieving this goal.

The theoretical foundation of the Data Reuploading Classifier can be analogized to the Universal Approximation Theorem (UAT) in artificial neural networks (ANNs). By repeatedly reuploading data, a single quantum bit can approximate any continuous function, similar to how a neural network with sufficient neurons can approximate any function. This analogy highlights the potential of the Data Reuploading process to enhance the expressiveness of quantum models, much like the way ANNs leverage multiple layers and neurons to represent complex functions.

The quantum circuit of the Data Reuploading Classifier usually contains several parts below:
\begin{enumerate}
    \item \textbf{Data Encoding} 
    
    The input of classical data $\bs{x} \in \mathbb{R}^d$ is encoded into the quantum state via parameterized unitary rotations $SU(2)$, writen as $R(\bs{\omega} \circ \bs{x})$ where $\bs{\omega}$ are tunable parameters.
    
    \item \textbf{Layered Circuits} 
    
    The circuit is composed of multiple layers $L(i)$, each of which consists of two components:
    
    \begin{equation}
        L(i) = R(\bs{\theta}_i) \circ R(\bs{\omega}_i \circ \bs{x})
    \end{equation}
    where $\bs{\theta}_i$ are tunable parameters . 
\end{enumerate}
The visual illustration of Data Reuploading Model is presented below.

%数据重载模型，由 Pérez-Salinas 等人（2020）提出，最初的目的是实现使用最小量子资源的通用量子分类器。其核心理念在于针对单个量子比特的计算空间有限（只有两个自由度）的问题，模型在量子线路中多次“重上传”数据，可以近似复杂的分类函数。数据重上传绕过了量子计算中的不可克隆定理（不允许直接复制量子数据），通过在多层量子线路中重新引入经典数据实现这一目标。
%数据重上传模型的理论基础可以类比于人工神经网络的通用逼近定理（UAT）。通过多次重上传数据，单个量子比特可以逼近任意连续函数，类似于具有足够神经元的神经网络可以逼近任何函数。
%数据重上传模型的量子线路包含以下几个关键组件：
%1.数据编码：量子机器学习通过使用参数化的单比特旋转门将经典输入数据编码到量子态中：  
%2.Layer Gates：量子线路由多层组成，每一层包括两个部分：
%3.测量与代价函数：对量子比特的最终状态进行测量，从而计算代价函数。最常用的选择是基于保真度的代价函数，它用于量化输出状态与目标标签状态之间的重叠程度。

\begin{figure*}[!t]
    \centering
    \includegraphics{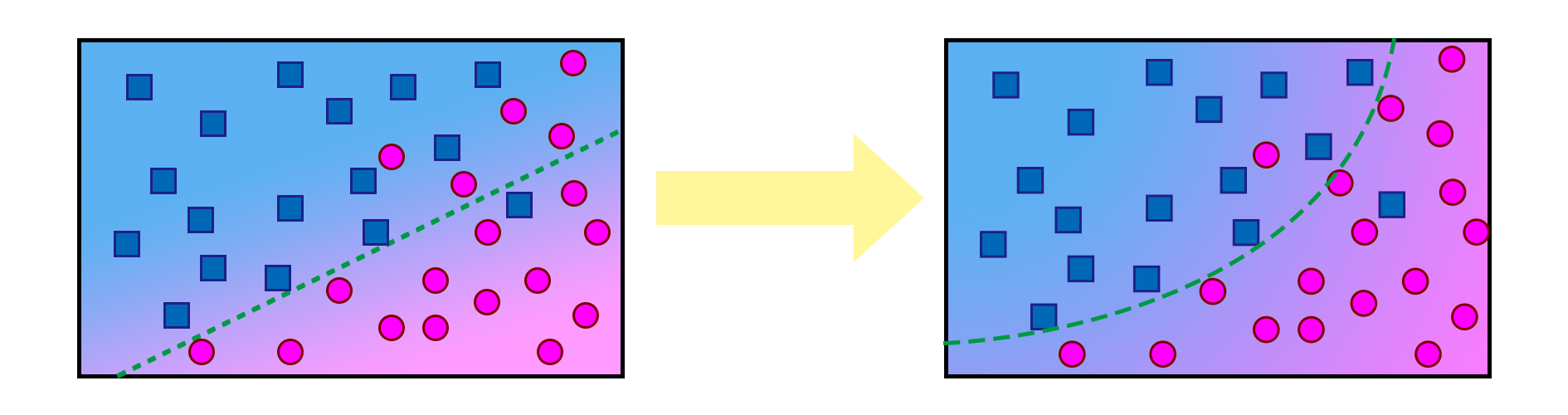}
    \caption{An illustration of classic machine learning: the input data is initially encoded linearly, and non-linear methods are applied later to enhance the model’s generalization capability. The yellow arrow means nonlinear methods such as kernel methods or activation functions.}
    \label{fig:enter-label}
\end{figure*}

\begin{figure*}[!t]
    \centering
    \includegraphics{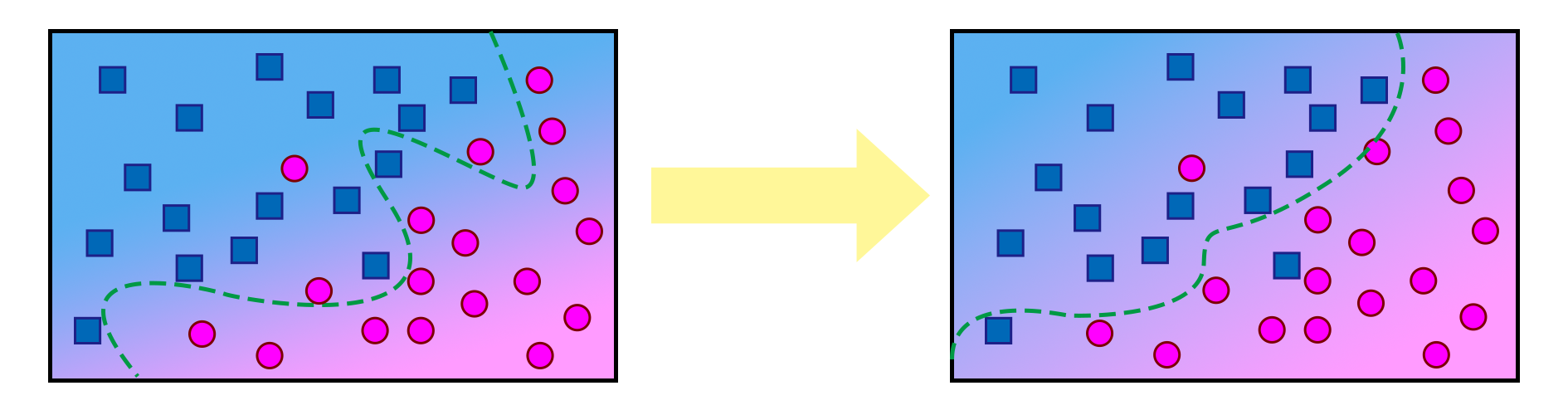}
    \caption{An illustration of MEDQ: Left shows the original quantum machine learning model where the input data is encoded non-linearly, while the right figure shows after integrating multiple embeddings, the MEDQ earned a better generalization capability.}
    \label{fig:enter-label}
\end{figure*}

\section{Circuit Architecture}
Despite the theoretical advantages of the data reloading model, its performance in practical applications remains suboptimal on certain datasets, particularly linearly separable datasets. It is quite surprising that a model capable of handling complex, non-linear datasets such as MNIST faces challenges when dealing with linearly separable datasets.

A closer examination of the structure of quantum machine learning models helps explain this phenomenon. Consider the analogy with binary logistic regression. Logistic regression is a generalization of linear regression, where an "activation function" is added to the linear model, enabling it to handle more complex problems. This approach mirrors the classical paradigm in machine learning, where input data is initially encoded into the model in a linear form to solve linearly separable classification problems. Subsequently, techniques such as kernel functions and activation functions are introduced to allow the model to adapt to more complex environments. In essence, the input data is initially encoded linearly, and non-linear methods are applied later to enhance the model’s generalization capability.

In quantum machine learning, this paradigm no longer holds. To enable quantum models to tackle classical machine learning tasks, data embedding is introduced as a solution. Classic input data is embedded into the quantum model through quantum gates, which use the data as parameters to rotate the qubits. The orthogonality of the rotation operator ensures that the classic data is inherently represented in a non-linear form upon embedding. This fundamental difference in structure between quantum and classical machine learning models results in distinct performance across different datasets.

This is especially true when the model has a limited number of layers, as the conditions required by the universal approximation theorem are not fully satisfied, thereby impacting the model's performance. In the data reuploading model, each instance of data embedding can be viewed as a neuron. While this structured design can effectively approximate the target function in some cases, it faces significant challenges in real-world applications. Specifically, with fewer layers, the embedding layer may fail to capture certain features present in the data.

As the number of layers in the model decreases, the impact of data embedding on the decision boundary becomes more pronounced, which directly leads to insufficient generalization capability when the model encounters complex data. This limitation of the data reuploading model is particularly evident when dealing with linearly separable datasets. This phenomenon highlights the model's inadequacy in feature extraction.

An important reason for this issue lies in the data reuploading model's over-reliance on a single data embedding strategy. Similar to artificial neural networks (ANNs), a single structure often fails to fully exploit the model's potential in complex datasets. A single data embedding typically limits the model's ability to extract multidimensional information from the data, thereby impacting its ability to model decision boundaries in classification tasks.

Simply changing the data embedding does not fully resolve the issue, which naturally leads to the idea of combining multiple data embeddings. While the concept is straightforward, its implementation presents a challenge. A linear combination of multiple data embeddings is an intuitive solution, and this approach is commonly used in classical machine learning. However, the structural differences between quantum machine learning models and traditional models complicate the implementation of such linear combinations. In quantum machine learning, the no-cloning theorem prevents data from being copied within quantum circuits. As a result, the linear combination of different data embeddings can only be achieved by increasing the number of qubits, which incurs an unacceptable computational cost. The data reuploading process, however, bypasses the no-cloning theorem by repeatedly uploading the data, creating a structure similar to that of neural networks, which provides insights into the combination of multiple data embeddings.

 Therefore, a new method is proposed to better generalize data by integrating multiple embeddings, which we call \textbf{M}ulti-\textbf{E}ncoding \textbf{D}ata reuploading \textbf{Q}uantum model (MEDQ).

To formalise the problem, let $\mathcal{X}$ be a set of input and $\mathcal{Y}$ be a set of output. The dataset $\mathcal{D}  = \{(\bs{x}_1,\bs{y}_1), (\bs{x}_2, \bs{y}_2), \cdots (\bs{x}_M, \bs{y}_M) \}$ is made of pairs of input data $\bs{x}_n \in \mathcal{X}$ and output data $\bs{y}_n \in \mathcal{Y}$. For simplicity, $\mathcal{X}  = \mathbb{R}^N$, $\mathcal{Y} = \{0, 1\}$, which is a binary classification task. 

Here's the mathematics form of MEDQ:

\begin{figure*}[!t]
    \centering
    \includegraphics[width = \linewidth]{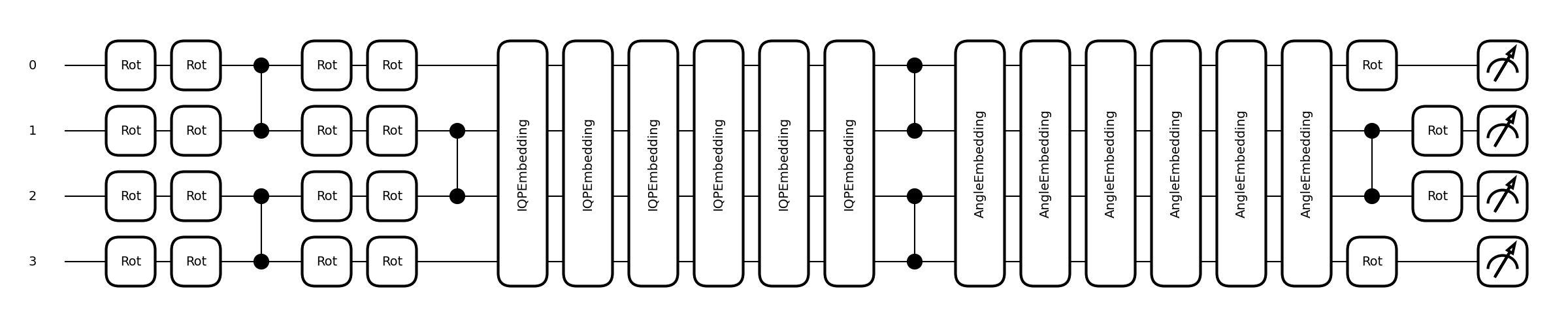}
    \caption{Visual illustration of Data Reuploading Model}
    \label{fig:data-reuplod}
\end{figure*}

\subsection{Data Embeddings}
    
    The quantum machine learning relies on the embedding process in order to solve classic machine learning problem. Each embedding process is written as $R(\bs{x})$. Below lie introductions of some common embeddings:

    \begin{enumerate}
        \item \textbf{Rot}
        
        Rot is the quantum gate used by the data reuploading model, which is writen as:
        \begin{equation}
        \begin{aligned}
            &R(\phi,\theta,\omega) = RZ(\omega)RY(\theta)RZ(\phi)\\
            =& \begin{bmatrix}
e^{-i(\phi+\omega)/2}\cos(\theta/2) & -e^{i(\phi-\omega)/2}\sin(\theta/2) \\
e^{-i(\phi-\omega)/2}\sin(\theta/2) & e^{i(\phi+\omega)/2}\cos(\theta/2)
\end{bmatrix}.
        \end{aligned}    
        \end{equation}
        
        \item  \textbf{QAOA Embedding}

        A single layer QAOA Embedding applies two circuits or “Hamiltonians”: The first encodes the features, and the second is a variational ansatz consisting of two-qubit ZZ interactions.The number of features has to be smaller or equal to the number of qubits. 

        \begin{equation}
            R(\bs{x}_1,\bs{x}_2) = [RY(\bs{x}_1)\quad RY(\bs{x}_2)] \circ ZZ(RX(\bs{x}_1), RX(\bs{x}_2))
        \end{equation}

        \item  \textbf{Angle Embedding}

        The Angle Embedding encodes every feature into the rotations of angles of the qubit. The length of features has to be smaller or equal to the number of qubits.
        \begin{equation}
            R(\bs{x}) = 
            \begin{cases}
                RX(\bs{x})\\
                RY(\bs{x})\\
                RZ(\bs{x})
            \end{cases}
        \end{equation}
    \end{enumerate}

\subsection{Reuploading Process}
    
    To combine multiple embeddings, as discussed earlier, while linear combination is an obvious and straightforward approach in machine learning, their computational cost becomes unacceptable due to the structural differences in quantum machine learning. Inspired by the quantum data reuploading model, we propose a solution where the same data is repeatedly uploaded to the same set of quantum bits through different embedding methods. This approach bypasses the no-cloning theorem and stores the information from different embeddings within the same set of qubits, essentially enabling the integration of different embedding information.

    The reuploading process is written as follows:
    \newline
    \newline
    \begin{equation}
        \begin{aligned}
        \ket{\bs{\psi}(\bs{\theta}, \bs{\omega}, \bs{x}
)} &= \Pi^n_{i=1}L(i)  \ket{0}\\
&= \Pi^n_{i=1}[R_i(\bs{\theta}_i)  R_i(\bs{\omega}_i \circ \bs{x})]\ket{0}
        \end{aligned}
    \end{equation}
    \newline
    \newline

    where $L(i) = R_i(\bs{\theta}_i)  R_i(\bs{\omega}_i \circ \bs{x})$ means each layer has its own variational parameters$\bs{\theta}_i$ and variable parameters $\bs{\omega}_i$, both of which are trainable.

\subsection{Measurement Process}

In a quantum system, after measurement, the system "collapses" to a specific eigenstate, and the measurement result corresponds to the eigenvalue of that eigenstate. Typically, the observation is performed using an operator, written as $O$, the observable operator. The mathematics form of the quantum system can be expressed as follows:
\newline
\newline
\begin{equation}
\begin{aligned}
    f(\bs{x}) &= Tr[\bs{\rho_x}O] = Tr[\Phi(\bs{\rho_x})]\\
    &= Tr[\Phi(\ket{\bs{\psi}(\bs{\theta}, \bs{\omega}, \bs{x}
)}\bra{\bs{\psi}(\bs{\theta}, \bs{\omega}, \bs{x}
)})]
\end{aligned}
\end{equation}
\begin{figure*}[t]
    \centering
    \begin{minipage}{0.49\textwidth}
        \centering
        \includegraphics[width = 0.49\textwidth]{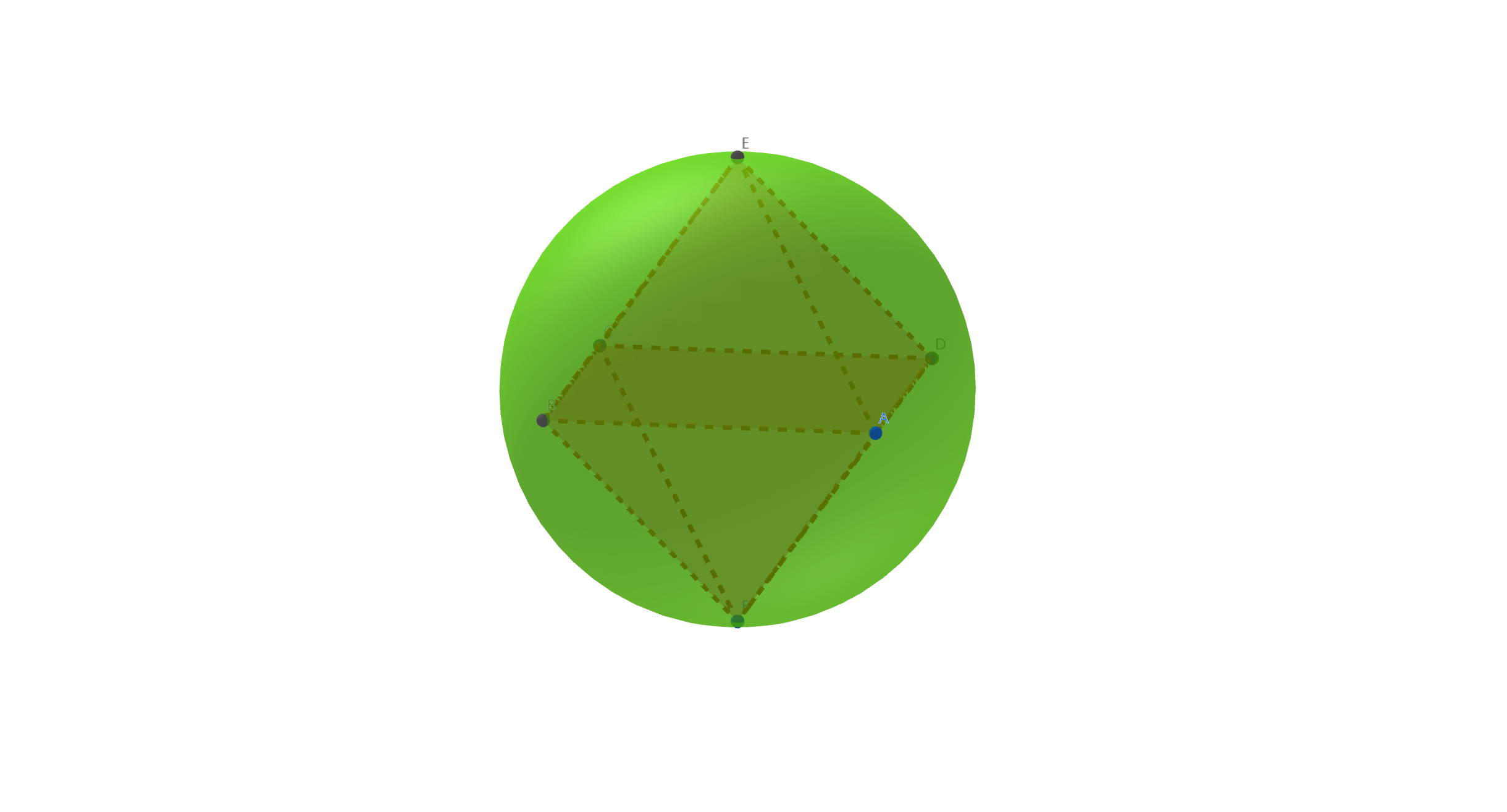}
    \end{minipage}
    \begin{minipage}{0.49\textwidth}
        \centering
        \includegraphics[width = 0.49\textwidth]{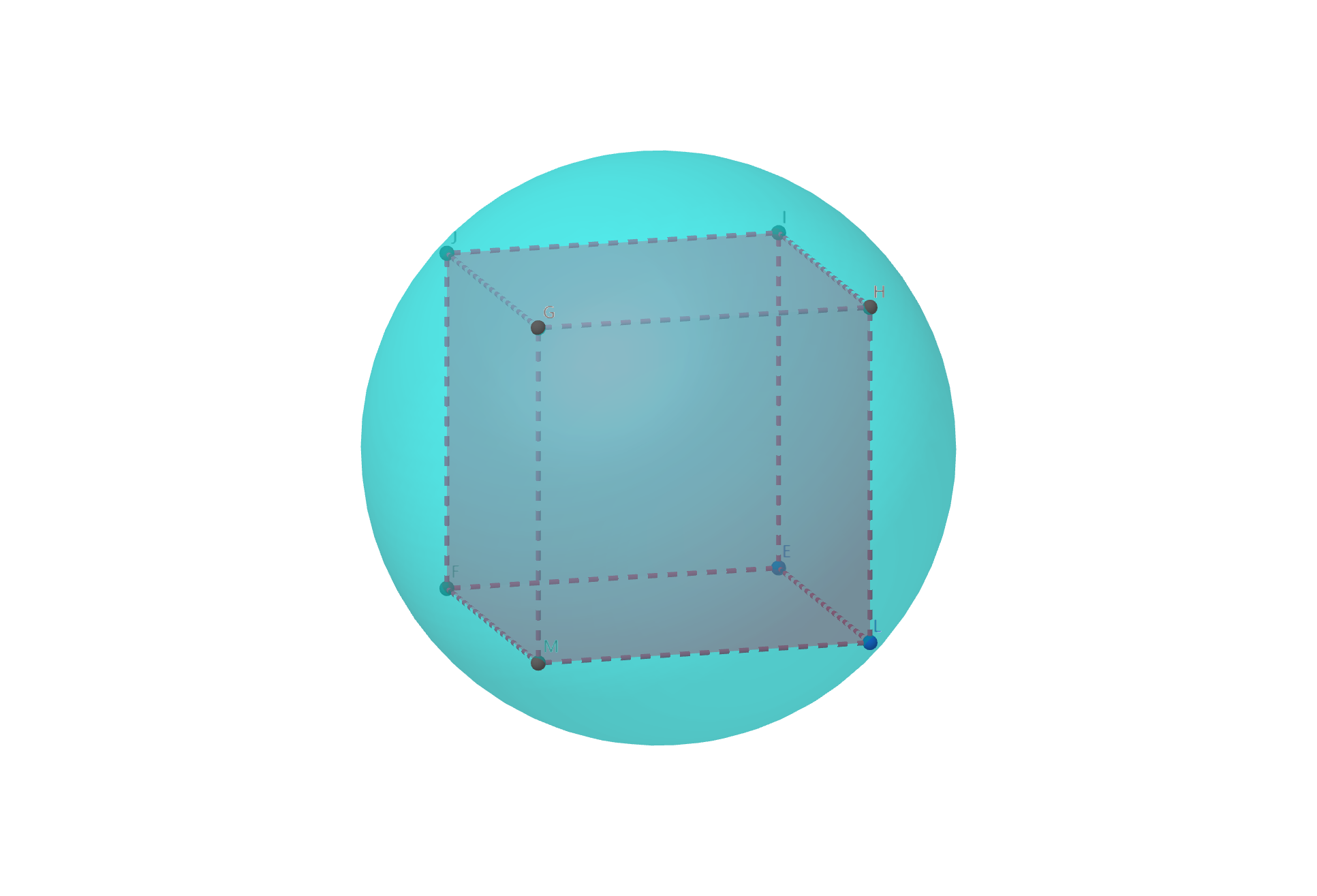}
    \end{minipage}
    \caption{Representation of the Bloch sphere, each point representing a class vector and single-qubit classifier will be trained to distributed the data points in one of these vertices}
\end{figure*}

\begin{table*}[b]
    \centering
    \begin{tabular}{cccccc}
     \toprule
     \textbf{Layer Num} &3 &4 &5 &6 &7  \\ \midrule
\textbf{MEDQ}&$\boldsymbol{0.9533}$&$\boldsymbol{0.9633}$&$\boldsymbol{0.9633}$&$\boldsymbol{0.9567}$&$\boldsymbol{0.98}$\\ \midrule
     \textbf{Data Reuploading} &0.7508 &0.8825 &0.9592 &0.9342 &0.9692\\ \midrule
     \textbf{Circuit Centric} &\multicolumn{5}{c}{0.5967}\\ \midrule
     \textbf{IQP Variational} &\multicolumn{5}{c}{0.8933}\\ \midrule
     \textbf{Quantum Metric} &\multicolumn{5}{c}{0.6333}\\ \midrule
     \textbf{Tree Tensor} &\multicolumn{5}{c}{0.5433}\\ \bottomrule
\end{tabular}
    \caption{Linear Separable - 10d}
    \label{tab:my_label}
\end{table*}

\subsection{Training Process}

The loss function chosen for MEDQ is the weighted quantum state fidelity loss function. Quantum state fidelity is an important metric for measuring the similarity between two quantum states. 
\begin{equation}
    F_c(\bs{\theta},\bs{\omega},\bs{x}) = |\bra{\bs{\psi}_c}\ket{\bs{\psi}(\bs{\theta},\bs{\omega},\bs{x})}|^2
\end{equation}

The weighted quantum state fidelity loss function further considers the relative importance of different quantum states in the loss function by introducing a weighting factor. In quantum neural networks, the target quantum state is typically defined as the quantum representation of the labels or some other known quantum state used as the training target. For instance, in quantum classification, the quantum state output by the model is compared with the target quantum state corresponding to the label, and the optimization goal is to minimize the fidelity loss between them.

\begin{equation}
    \chi^2_{wf}(\bs{\theta},\bs{\omega}, \bs{\alpha}) = \frac{1}{2}\sum^M_{\mu=1}\left(\sum^{\mathcal{C}}_{c=1}(\alpha_cF_c(\bs{\theta},\bs{\omega},\bs{x}_\mu) - Y_c(\bs{x}_\mu) )^2 \right)
\end{equation}
Here $Y_c$ represents the expected quantum state fidelity in the case of a successful classification.

\begin{table*}[b]
    \centering
    \begin{tabular}[width = 0.6*\linewidth]{ccccccccc}
     \toprule
     \textbf{Layer Num} &3 &4 &5 &6 &7 &8 &9 &10  \\ \midrule
     
        \textbf{ MEDQ} &0.7258	&0.7258	&$\boldsymbol{0.7332}$	&$\boldsymbol{0.7332}$	&$\boldsymbol{0.7288}$	&$\boldsymbol{0.7332}$	&0.7288	&$\boldsymbol{0.7332}$\\ \midrule
        \textbf{Data Reuploading} &$\boldsymbol{0.7309}$	&$\boldsymbol{0.7309}$	&0.7309	&0.7309	&0.7189	&0.7309	&$\boldsymbol{0.7309}$	&0.7309\\ \midrule
    \textbf{Circuit Centric} &\multicolumn{8}{c}{0.6451}\\ \midrule
     \textbf{IQP Variational} &\multicolumn{8}{c}{0.7210}\\ \midrule
     \textbf{Quantum Metric} &\multicolumn{8}{c}{0.6981}\\ \midrule
     \textbf{Tree Tensor} &\multicolumn{8}{c}{0.4818}\\ \bottomrule
    \end{tabular}
    \caption{MNIST - 3d}
    \label{tab:my_label}
\end{table*}

\begin{table*}[b]
    \centering
    \begin{tabular}[width = 0.6*\linewidth]{ccccccccc}
     \toprule
    \textbf{ Layer Num }&3 &4 &5 &6 &7 &8 &9 &10  \\ \midrule
     
         \textbf{MEDQ} &$\boldsymbol{0.9359}$	&$\boldsymbol{0.9359}$	&$\boldsymbol{0.9366}$	&$\boldsymbol{0.9359}$	&$\boldsymbol{0.9359}$	&$\boldsymbol{0.9366}$	&$\boldsymbol{0.9366}$	&$\boldsymbol{0.9366}$
    \\ \midrule
    \textbf{Data Reuploading} &0.9338	&0.9338	&0.9338	&0.9293	&0.9338	&0.9338	&0.9313	&0.9338
    \\ \midrule
    \textbf{Circuit Centric} &\multicolumn{8}{c}{0.7861}\\ \hline
     \textbf{IQP Variational} &\multicolumn{8}{c}{0.8732}\\ \hline
     \textbf{Quantum Metric} &\multicolumn{8}{c}{0.8699}\\ \hline
     \textbf{Tree Tensor} &\multicolumn{8}{c}{0.5291}\\ \hline
    \end{tabular}
    
    \caption{MNIST - 5d}
    \label{tab:my_label}
\end{table*}
%首先在这里插入一幅大体的示意图，然后再对问题的产生做进行更细致地阐述

%然而，尽管数据重载模型在理论上具备一定的优势，实际应用中其表现却在某些数据集上，尤其是线性可分数据集，仍然不尽如人意。这是非常令人震惊的，一个能够处理形如MNIST这样的非线性复杂数据集的模型会在处理线性的数据集时遭遇障碍。

%但是其实如果我们仔细考虑量子机器学习模型的结构，就能够理解这个看似不符合常理的现象。让我们以二元logistic回归分类作一个类比。logistic回归是一个广义上的线性回归，它通过在线性回归的基础上添加了一层“激活函数”，从而使模型可以应对更加复杂的环境。这是经典数据中机器学习的范式：模型的数据通常以线性的形式载入到模型之中解决线性可分的分类问题，之后引入了形如核函数，激活函数等手段，使得模型可以适应更加复杂的环境。也就是模型载入的数据是以线性方式存在，之后通过非线性的手段增强模型的拟合能力。

%而在量子机器学习，这个范式却不再适用：为了能够使用量子模型处理经典数据的机器学习任务，数据编码作为解决方法被引入。它通过量子门将经典的数据作为对量子比特进行旋转操作的参数引入到量子模型中。旋转算子的正交性使得经典数据在载入时就以非线性的形式存在，这也使得量子机器学习模型和经典数据中的机器学习模型从基本的结构上就产生了差别，进而造就了两种模型在不同数据集表现上的差异。

%尤其是在模型的层数较少时，万能逼近定理的条件无法得到充分满足，从而影响了模型的表现。在数据重载模型中，每一次数据编码都可以被视为一个神经元，而这种结构化的设计虽在某些情况下能够有效地逼近目标函数，但在实际应用中却面临着一定的挑战。当层数较少时，数据编码的效果未必能充分捕捉到数据中的复杂特征。

%更为重要的是，随着模型层数的减少，数据编码在决策边界上的影响变得更加突出，这直接导致了模型在面对复杂数据时的拟合能力不足。数据重载模型的这种局限性，尤其是在面对线性可分数据集时表现得尤为明显。这一现象反映了模型在特征提取方面的不足。

%这一问题的一个重要原因在于数据重载模型对于单一数据编码的过度依赖。与传统的人工神经网络（ANNs）类似，单一结构往往无法充分发挥模型在复杂数据集中的表现潜力。单一的数据编码往往会限制模型从数据中提取多维度信息的能力，从而影响其在分类任务时对决策边界的建模。

%这个问题并不是能够通过更换一种数据编码就能够得到解决的。因此，我们提出了一个能够综合多种数据编码从而拥有更强拟合能力的模型：多编码重上传量子模型(Multi-Encoding Datareuploading Quantum model)

\section{Experiments}

It can be seen that the method we propose is essentially an extension of the data reuploading model, offering strong versatility. We can freely select the embedding strategies and their arrangements as needed. To facilitate a better comparison with the data reuploading model, we have designed the structure above.

The model consists of 3n layers, where the first n layers are the embeddings used by data reuploading model, while the second n layers are QAOA Embeddings and the last n layers are Angle Embeddings.

This architecture, while ensuring the applicability of the universal approximation theorem to quantum circuits, allows for more targeted optimization of the encoding under the condition of having the same parameters, allowing for meaningful comparisons.

In our experiments, we selected both linearly separable datasets and the MNIST dataset for evaluation. We compared our model with several others, including the Data Reuploading Model, Circuit Centric Classifier, IQP Variational Classifier, and Quantum Metric Learner. The results of Circuit Centric Classifier, IQP Variational Classifier, and Quantum Metric Learner are chosen in the benchmarks\cite{bowles2024betterclassicalsubtleart}

Our model demonstrated performance no worse than that of state-of-the-art methods, showcasing superior generalization capability across a wider range of datasets compared to baseline. The detailed results are shown below.

\subsection{Linear Separable}
The datasets consist of inputs randomly sampled from a d-dimensional hypercube, divided into two classes by the hyperplane orthogonal to the $(1, ...., 1)^T$ vector with a small data-free margin. It is easy to understand and clearly defined. More importantly, previous studies have discovered that quantum machine learning methods struggle with the linear separable benchmarks, which indicates the limitation in the generalization capability of current quantum machine learning models. 

We conduct experiments on datasets with 10, 12, and 14 dimensions. In particular, for the two structurally similar models, we perform experiments per number of layers. For each number of layers, we perform a grid search over the remaining hyperparameters to obtain the hyperparameters group that minimizes the training error. Each configuration is tested five times and the average performance is reported. This approach ensures a fair and accurate evaluation of the models' performance in practical usage scenarios. and we attain a satisfying result that our model outperforms the data reuploading model by 20\% at most.  

\begin{table*}[h]
    \centering
    \begin{tabular}[width=0.48*\linewidth]{cccccc}
     \toprule
    \textbf{ Layer Num} &3 &4 &5 &6 &7  \\ \midrule
     
     \textbf{MEDQ}
&$\bs{0.8967}$	&$\boldsymbol{	0.9633}$	&$\boldsymbol{	0.96	}$	&$\boldsymbol{0.9767	}$	&$\boldsymbol{0.9833}$\\ \midrule
\textbf{Data Reuploading} &0.8033	&0.9033	&0.9133	&0.9533	&0.9167\\ \midrule
\textbf{Circuit Centric} &\multicolumn{5}{c}{0.61}\\ \midrule
     \textbf{IQP Variational} &\multicolumn{5}{c}{0.61}\\ \midrule
     \textbf{Tree Tensor} &\multicolumn{5}{c}{0.56}\\ \bottomrule
\end{tabular}
    \caption{Linear Separable - 12d}
    \label{tab:my_label}
\end{table*}

We can evaluate the experimental results from two perspectives: in terms of optimal performance, it is evident that the MEDQ model outperforms the current state-of-the-art (SOTA) methods. This demonstrates the significant improvement in the model’s generalization capability achieved by our proposed framework.

When comparing the optimal performance at each number of layers, we focus on the two models that require multiple data reuploads: MEDQ and the Data Reuploading Model. From this comparison, we can draw the conclusion that for the same number of layers, the performance of MEDQ consistently exceeds that of the Data Reuploading Model.

Notably, MEDQ achieves optimal performance with fewer layers than the Data Reuploading Model. This is particularly advantageous in practical applications, as it indicates that MEDQ requires fewer parameters to achieve the same task.

Both of these experimental findings strongly validate the effectiveness of our framework and its robust generalization capability.

\begin{table*}[h]
    \centering
    \begin{tabular}[width=0.48*\linewidth]{cccccc}
     \toprule
    \textbf{ Layer Num} &3 &4 &5 &6 &7  \\ \midrule
     
     \textbf{MEDQ}&$\bs{0.8767}$	&$\boldsymbol{	0.8633}$	&$\boldsymbol{	0.9267	}$	&0.7433	&0.5833\\ \midrule
     \textbf{Data Reuploading} &0.7333	&0.79	&0.7333	&$\boldsymbol{0.9267}$	&$\bs{0.83}$\\ \midrule
\textbf{Circuit Centric} &\multicolumn{5}{c}{0.5267}\\ \midrule
     \textbf{IQP Variational} &\multicolumn{5}{c}{0.6}\\ \midrule
     \textbf{Tree Tensor} &\multicolumn{5}{c}{0.5333}\\ \bottomrule
\end{tabular}
    \caption{Linear Separable - 14d}
    \label{tab:Linear14d}
\end{table*}

\subsection{MNIST}
MNIST is a classic machine learning dataset widely used for image classification and pattern recognition tasks, particularly in the fields of deep learning and computer vision. It consists of a large collection of handwritten digit images and is commonly used to test and compare the performance of various machine learning algorithms. We use Principal Component Analysis (PCA) to reduce the dimensions for the quantum machine learning models.  

We conduct experiments on datasets with 3 and 5 dimensions, where MEDQ demonstrates outstanding performance in both cases. This indicates that MEDQ inherits the excellent generalization capability of the data reuploading model across various datasets.

\section{Conclusion \& Discussion}
\subsection{Discussion}
From the table \ref{tab:Linear14d}, there is a decline in performance as the number of layers increases. While the training accuracy remains high, the overall performance deteriorates, a phenomenon that we attribute to overfitting. Future research could focus on addressing this issue by incorporating regularization techniques, or Early Stopping.

Besides, though the MEDQ model has demonstrated its strong generalization capability through experiments, we have provided a reasonable explanation for this, and its generalization ability is also theoretically supported, further theoretical validation is required to establish its advantage over the data reuploading model. Additionally, the reliability and security of the model will need to be examined through subsequent theoretical studies.

Moreover, the framework proposed in this paper highlights the potential of the MEDQ model. However, the selection of specific embeddings, their arrangement, and the tuning of hyperparameters such as learning rate remain open problems. Further research in these areas will provide valuable guidance for future applications.

\subsection{Conclusion}
In this paper, we proposed a novel Multi-Encoding Data reuploading Quantum model (MEDQ), which integrates multiple quantum data embedding strategies to enhance the generalization capability of quantum machine learning models. Our experimental results demonstrated that MEDQ outperforms existing state-of-the-art methods, showing superior generalization ability across a wide range of datasets.

The proposed MEDQ framework not only enhances model generalization but also offers a flexible approach for encoding classical data in quantum machine learning tasks. This advancement has significant implications for the development of quantum machine learning models capable of addressing increasingly complex datasets and applications.

In conclusion, the MEDQ model presents a significant step forward in the field of quantum machine learning, providing a powerful tool for handling complex and diverse datasets. As quantum hardware continues to evolve, we expect this framework to play a pivotal role in advancing both theoretical and practical applications of quantum machine learning.

\newpage
\newpage

\bibliographystyle{named}
\bibliography{ijcai25}

\end{document}